# Emergence of a symmetry-broken Chern insulator near a moiré Kondo breakdown

Wanghao Tian[1†], Bowen Shen[2,3†], Lizhong Li[2,3], Mingjie Zhang[4], Feng Liu[1], Chushan Li[1,5], Yaotian Liu[4], Fan Xu[1,5], Kenji Watanabe[6], Takashi Taniguchi[7], Peiling Li[4], Li Lu[4,8], Yang Xu[4], Shengwei Jiang[1,5*], Tingxin Li[1,5,8*], Jie Shan[2,3,9*], Kin Fai Mak[2,3,9*]

[1]State Key Laboratory of Micro-nano Engineering Science, Key Laboratory of Artificial Structures and Quantum Control (Ministry of Education), School of Physics and Astronomy, Shanghai Jiao Tong University, Shanghai, China
[2]School of Applied and Engineering Physics and Laboratory of Atomic and Solid State Physics, Cornell University, Ithaca, NY, USA
[3]Kavli Institute at Cornell for Nanoscale Science, Ithaca, NY, USA
[4]Beijing National Laboratory for Condensed Matter Physics and Institute of Physics, Chinese Academy of Sciences, Beijing, China
[5]Tsung-Dao Lee Institute, Shanghai Jiao Tong University, Shanghai, China
[6]Research Center for Electronic and Optical Materials, National Institute for Materials Science, 1-1 Namiki, Tsukuba, Japan
[7]Research Center for Materials Nanoarchitectonics, National Institute for Materials Science, 1-1 Namiki, Tsukuba, Japan
[8]Hefei National Laboratory, Hefei, China
[9]Max Planck Institute for the Structure and Dynamics of Matter, Hamburg, Germany

[†]These authors contribute equally to this work.
[*]Emails: swjiang@sjtu.edu.cn, txli89@sjtu.edu.cn, jie.shan@mpsd.mpg.de, kin-fai.mak@mpsd.mpg.de

**Abstract**
Moiré semiconductors built on angle-aligned transition metal dichalcogenide (TMD) heterobilayers provide a physical realization of the Kondo lattice model, in which one TMD layer is prepared in a Mott insulating state supporting a lattice of local magnetic moments and the other layer in a metallic state supporting itinerant carriers. The artificial Kondo lattice enables the exploration of exotic states of matter near a continuously tunable Kondo breakdown. Here we report the emergence of a symmetry-broken Chern insulator at a moiré hole filling factor 4/3 in angle-aligned $MoTe_2/WSe_2$ moiré bilayers, which realize a chiral Kondo lattice. The symmetry-broken Chern insulator, which exhibits integer quantized Hall conductance at a fractional moiré filling, breaks the translational symmetry of the lattice spontaneously; it also appears only near a magnetic field-induced Kondo breakdown in the mixed-valence regime of the material. We further demonstrate that the magnetic field required to induce the Kondo breakdown and to stabilize the symmetry-broken Chern insulator is twist angle dependent. The results present new opportunities for exploring the subtle interplay between topology and Kondo interactions in moiré semiconductors.

**Main**

The Kondo lattice, a lattice of localized magnetic moments coupled via exchange interactions to a sea of itinerant electrons, is a prototypical strongly correlated system[1,2]. Driven by the strong Coulomb repulsion and the Kondo interaction, the system can host a suite of quantum phases of matter, such as heavy Fermi liquids[3,4], Kondo insulators[5,6], quantum critical non-Fermi liquids[7-9] and unconventional superconductivity[10-12]. Whereas Kondo lattice physics has been extensively studied in lanthanide and actinide intermetallic compounds[13-16], the emergence of two-dimensional moiré materials has presented a new platform to study the Kondo lattice problem in a continuously tunable manner[17-25]. In particular, varying the moiré lattice filling factor and/or an electric field perpendicular to the sample plane in these materials can tune the effective Kondo interaction and induce a breakdown of the Kondo singlets, near where exotic states of matter could emerge.

In this work, we report the experimental observation of a symmetry-broken Chern insulator at a moiré hole filling factor $\nu = 4/3$ and near a magnetic field-induced Kondo breakdown in dual-gated devices of angle-aligned $MoTe_2/WSe_2$ moiré bilayers (Fig. 1a). The top and bottom gates in these devices allow independent tuning of $\nu$ and the vertical electric field $E$. (See Methods for details on device fabrication and electrical transport measurements.) Angle-aligned $MoTe_2/WSe_2$ bilayers[18-22,26-28] form a hexagonal moiré lattice featuring three high-symmetry stacking sites: MM, MX, and XX, where M represents Mo or W and X denotes Te or Se (Fig. 1b). The 7% lattice mismatch between the Mo- and W-layer gives a moiré period $a_M \approx 4.8$nm and a moiré density $n_M \approx 4.5 \times 10^{12}$ cm$^{-2}$. The Wannier orbitals of the Mo- and W-layer are centered at the two sublattice sites of the lattice[29]; the sublattice potential and thus the energy separation between the Mo- and W-moiré bands are continuously tunable by the electric field $E$.

**Electrostatics phase diagram**

Fig. 1c shows a schematic phase diagram of the material, which can be separated into five different regions (I-V) with boundaries marked by dashed lines, as reported in previous studies[18-22]. The corresponding band alignment between the two layers is shown in Fig. 1d. In region I, only the Mo-layer is hole-doped with a hole filling factor $\nu = \nu_{Mo} + \nu_W = \nu_{Mo} + 0$ (here $\nu_{Mo}$ and $\nu_W$ denote the hole filling factor in the Mo- and W-layer, respectively). Insulating states were reported at $\nu_{Mo} = 1$ and 2, corresponding to a Mott and a band insulator, respectively[18,22,27]. As a result of the strong on-site Coulomb repulsion ($U$) compared to the bandwidth in the Mo-layer, the topmost Mo-moiré band is split into the lower and upper Hubbard bands. Similarly, only the topmost W-moiré band is hole-doped in region II, i.e. $\nu = 0 + \nu_W$. Because of the substantially larger bandwidth in the W-layer, no insulating state can be identified at $\nu_W = 1$, demonstrating that holes in the W-layer are much less correlated than those in the Mo-layer[22].

Region IV corresponds to the Kondo lattice region. Here the Mo-layer is kept at the Mott insulating state (i.e. $\nu_{Mo} = 1$) supporting a triangular lattice of local magnetic moments; the Mott gap size is proportional to the electric field span in this region[18]. The W-layer hosts a sea of itinerant holes with tunable density $\nu_W$; the total hole filling is $\nu = 1 + \nu_W$. The low-energy physics in this region can be mapped to a chiral Kondo lattice model carrying an exchange interaction ($J$) between the local moments in the Mo-layer, an intralayer hopping term ($t$) for the itinerant holes in the W-layer and a Kondo exchange interaction ($J_K$) between the spins in the Mo- and W-layer; $J_K$ further carries a *p*-wave chiral form factor that gives rise to a topologically nontrivial Kondo hybridization gap[23]. The emergence of chiral Kondo lattice physics in this region is supported by the observation[18,21,22] of 1) a tunable heavy Fermi liquid, 2) a Kondo breakdown (induced by magnetic field or by variation in $v_W$) from a heavy Fermi liquid with density $1 + v_W$ to an itinerant Fermi liquid with density $v_W$ decoupled from the local moments, 3) a Chern metal near the Kondo breakdown, and 4) a topological Kondo insulator at $v = 1 + 1$.

Outside the Kondo lattice region, the material enters the mixed-valence regions (III and V). The upper (lower) part corresponds to $v = v_{Mo} + v_W$ with $v_{Mo} < 1$ in region III ($v_{Mo} > 1$ in region V). The hole Fermi level cuts through both the W-moiré band and the Mo-Hubbard band here; only the singly occupied sites in the Mo-layer carry local magnetic moments; the local moment density is strongly fluctuating. We will focus on region V in this study, where the Fermi level cuts through the W-moiré band and the Mo-upper Hubbard band.

**Emergence of a symmetry-broken Chern insulator at $v = 4/3$**

Figures 1e and 1f show the longitudinal resistance ($R_{xx}$) of Device I (with nearly perfect angle-alignment, i.e. $a_M \approx 4.8$nm and $n_M \approx 4.6 \times 10^{12}$cm$^{-2}$) as a function of $v$ and $E$ at a perpendicular magnetic field $B = 0$T and 4.4T, respectively. The corresponding Hall resistance ($R_{xy}$) map at $B = 4.4$T is shown in Fig. 1g. Unless otherwise specified, the measurement temperature for Device I is at a lattice temperature $T = 20$mK. The grey-shaded areas are the inaccessible regions in the phase diagram due to limited gate voltage range and/or large contact/sample resistance. The dashed lines mark the phase boundaries for the regions I, III and IV as discussed above.

As shown in Fig. 1e, an electric-field-induced phase transition from a Mott insulator to a quantum anomalous Hall (QAH) insulator and then to an antiferromagnetic insulator was observed at $v = 1$ and $B = 0$T, in agreement with previous studies[22,27,30]. At $B = 4.4$T (Fig. 1f), vertical stripes of Shubnikov-de Haas (SdH) oscillations are observed in region IV; they help to identify the phase boundaries marked by black dashed lines. The oscillations originate from the Landau levels in the W-layer after a magnetic field-induced Kondo breakdown. Moreover, $R_{xy}(< 0)$ corresponds to a hole Fermi surface with density $v_W$ in most of the accessible area within region IV. These results demonstrate the emergence of an itinerant hole gas in the W-layer after the Kondo breakdown; the hole gas is decoupled from the Mott insulator in the Mo-layer. These observations are fully consistent with earlier reports on the same material[18-22].

The most striking observation is the emergence of a $R_{xx}$ dip (surrounded by an enhanced $R_{xx}$) and a $R_{xy}$ hot spot at $v \approx 4/3$ under $B = 4.4$T in region V (the lower part of the mixed-valence region). The state spans a finite electric field range in the region. The $R_{xy}$ hot spot is also located near a boundary, where $R_{xy}$ changes sign (dotted line in Fig. 1g). Note that the $R_{xx}$ dip and $R_{xy}$ hot spot in region III correspond to the QAH state at $v = 1$.

We further investigate the nature of the $v = 4/3$ state. Figures 2a and 2b show $R_{xx}$ and $R_{xy}$, respectively, as a function of $v$ and $B$ at $E = 0.575$V/nm. The electric field is chosen near the center of the $v = 4/3$ state as shown in Fig. 1g. The $R_{xx}$ dip and $R_{xy}$ hot spot near $v = 4/3$ start to appear at $B \approx 2$T. A sign change in $R_{xy}$ is also observed at a critical magnetic field ($B_C$); $B_C$ increases monotonically with $v$; the $v = 4/3$ state disappears abruptly at $B_C \approx 4.7$T for Device I. A set of Landau fans originating from $v = 1$ also appears for $B > B_C$.

Figure 2c shows a representative line cut at $B = 4.4$T. A quantized $R_{xy}$ at $\frac{h}{e^2}$, accompanied by a $R_{xx}$ dip (down to about 200Ω), is observed at $v = 4/3$. We also show in Fig. 2d the magnetic field dependence of $R_{xx}$ and $R_{xy}$ along the dashed line in Fig. 2a,b. $R_{xy}$ increases quickly above $B \approx 2$T, becomes nearly quantized at $\frac{h}{e^2}$ for $3 \lesssim B \lesssim 5$T, and abruptly decreases in magnitude and changes sign at $B_C$. Accompanied by this is the appearance of $R_{xx}$ peaks at $B \approx 2$T and 5T and a vanishing $R_{xx}$ in between.

The integer quantized $R_{xy} = \frac{h}{e^2}$ and the vanishing $R_{xx}$ at a fractional filling $v = 4/3$ demonstrate the emergence of a symmetry-broken Chern insulating state, or equivalently, a topological charge density wave state[31-37]. The conclusion is further supported by the dispersion of the $v = 4/3$ state with magnetic field. The slope of the dispersion follows the Streda formula $n_M \frac{dv}{dB} = C\frac{e}{h}$ with a Chern number close to $C = -1$ (dashed lines). The symmetry-broken Chern insulator appears only in a narrow field window of $2 \lesssim B \lesssim 5$T. It is not an integer quantum Hall state as it is not a part of the Landau fan that appears at $B \gtrsim B_C$; the abrupt disappearance of the state at $B_C$ is also inconsistent with a quantum Hall state. The energy gap of the symmetry-broken Chern insulator can be estimated by the temperature dependence of $R_{xx}$ and $R_{xy}$ (at $B = 4.4$T) shown in Fig. 2e. The quantized anomalous Hall transport disappears at a temperature around 1K (Fig. 2e and Supplementary Fig. 1). A thermal activation fit (Fig. 2f) to the $R_{xx}$ temperature dependence gives an energy gap about 0.25meV (or 3K).

**Relationship between the $v = 4/3$ Chern insulator and Kondo breakdown**
To demonstrate the connection between the $v = 4/3$ symmetry-broken Chern insulator and Kondo physics, we first show the emergence of heavy fermion physics through temperature and magnetic field dependent transport studies in the mixed-valence region V (Supplementary Fig. 2). In this region, the Fermi level intersects both the W-moiré band and the Mo-upper Hubbard band (Fig. 1d). Consistent with earlier results in the Kondo lattice regime[18,22], coherent heavy fermion transport develops at temperatures below the Kondo coherence temperature $T^*$; a temperature dependence, $R_{xx} = R_0 + A \times T^2$, is observed in the low-temperature limit, from which the residual resistance $R_0$ and the Kadowaki-Woods coefficient $A$ (proportional to the square of the quasiparticle effective mass) can be extracted. For $B > B_C$, the Zeeman energy outcompetes the effective Kondo interaction, leading to a breakdown of the Kondo singlets and the emergence of SdH oscillations due to a drop in the effective mass (Ref. [9,18,22,38]). This is accompanied by a sign change in $R_{xy}$ and a jump in the Hall density, reflecting the reconstruction from a large electron-like Fermi surface to a small hole-like Fermi surface (Supplementary Fig. 3). The effective Kondo exchange interaction, as reflected by $T^*$, $A^{-0.5}$ and $B_C$, also increases with $\nu$.

With heavy fermion physics established, we show in Fig. 3a-h the evolution of $R_{xy}$ maps versus $\nu$ and $E$ with increasing magnetic field. We mark the Kondo breakdown boundary, where $R_{xy}$ changes sign, by dotted lines. The Kondo breakdown boundary extends smoothly from region IV (Kondo lattice region) to region V (mixed-valence region), and moves to higher fillings as the field increases. Interestingly, the $R_{xy}$ hot spot around $\nu = 4/3$ starts to show up as the Kondo breakdown boundary approaches $\nu = 4/3$ in the mixed-valence region. Maximum $R_{xy}$ is observed when the boundary is at a filling slightly less than $\nu = 4/3$, where the $R_{xy}$ hot spot at $\nu = 4/3$ nearly spans the entire mixed-valence region (Fig. 3c-e). The $R_{xy}$ hot spot immediately disappears when the boundary passes $\nu = 4/3$ (Fig. 3f,g). The abrupt disappearance of the $\nu = 4/3$ state for $B > B_C$ can also be seen in Fig. 2d. The data show that the $\nu = 4/3$ symmetry-broken Chern insulator becomes the most stable right before the Kondo breakdown, where fluctuation effects are enhanced as the effective Kondo interaction becomes critically weak. The appearance of the $\nu = 4/3$ state only in region V but not in region IV also implies the importance of local-moment fluctuations that are inherent to the mixed-valence region.

**Twist angle dependence**
We also studied another device (Device II) with a finite twist angle at 1.5°, which corresponds to $n_M \approx 5.1 \times 10^{12}$cm$^{-2}$ and $a_M \approx 4.5$nm. Figures 4a and 4b show $R_{xx}$ and $R_{xy}$, respectively, as functions of $\nu$ and $B$ at $T = 300$mK and at a fixed electric field $E = 0.605$V/nm that cuts through region V at approximately the same location as Device I. Compared to Device I, the $\nu = 4/3$ Chern insulator appears at higher magnetic fields $5 \lesssim B \lesssim 7$T; the critical field $B_C$ for Kondo breakdown is also higher. The shift to higher fields could be explained by the higher moiré density $n_M$, which is expected to enhance the effective Kondo interaction and the Kondo breakdown critical field[18]. The twist angle dependence of the magnetic field range that hosts the $\nu = 4/3$ state is shown in Supplementary Fig. 4 (for a total of four devices). Whereas the overall field scale increases with the twist angle, the field range decreases slightly.

The $R_{xx}$ and $R_{xy}$ maps versus $\nu$ and $E$ at two selected magnetic fields $B = 6.6$T and 10T are shown in Fig. 4c-f (see Supplementary Fig. 5 for data at higher fields). Similar to Device I, the $\nu = 4/3$ state appears only in the mixed-valence region (region V) and is stabilized only near the Kondo breakdown, whose boundary is marked by black dotted lines in the $R_{xy}$ maps. After the breakdown at $B = 10$T, two different sets of SdH oscillations crossing each other emerge in region V; the two sets correspond to the two different types of carriers in the system: decoupled holes in the Mo-layer and W-layer. (The former set is less clear in the $R_{xx}$ map due to the heavier effective mass in the Mo-layer.) These observations further illustrate that the $\nu = 4/3$ state is intimately linked to the Kondo breakdown, rather than being induced by a change in the band alignment in the two different regions (IV and V).

Whereas the overall behavior of Device II is consistent with that of Device I, no $R_{xx}$ dip is observed at the $R_{xy}$ hotspot in Device II, and the maximum $R_{xy}$ only reaches about 10kΩ. The absence of a quantized Hall response and the lack of a concurrent $R_{xx}$ dip show that the symmetry-broken Chern insulator at $v = 4/3$ is not fully developed in Device II. On the other hand, the QAH state at $v = 1$ still shows quantized $R_{xy}$ and vanishing $R_{xx}$ at $T \lesssim 2\,\mathrm{K}$, demonstrating the reasonably high device quality (Supplementary Fig. 6). These observations suggest that the stability of the $v = 4/3$ Chern insulator is highly sensitive to the twist angle, further underscoring the delicate interplay between charge ordering, topology, and the proximity to the Kondo breakdown.

**Discussion and outlook**

The observation of a symmetry-broken Chern insulator at $v = 4/3$ near a magnetic field-induced Kondo breakdown in the mixed-valence region raises many interesting questions for further theoretical and experimental studies. First, although the state emerges at a fractional filling factor, it is an integer rather than a fractional Chern insulator[33,40-44]. The observation suggests a rather nonuniform distribution of the Berry curvature for the moiré bands in angle-aligned MoTe$_2$/WSe$_2$; the nonuniform distribution is likely connected to the dispersive moiré band in the W-layer[39]. Second, the emergence of the Chern insulator near the field-induced Kondo breakdown highlights the importance of the chiral Kondo interaction in angle-aligned MoTe$_2$/WSe$_2$, as pointed out by theoretical studies[23,24]. The chiral Kondo interaction introduces a topological Kondo hybridization gap between the dispersive W-moiré band and the flat Mo-Hubbard band, resulting in topologically protected edge states inside the gap. However, exactly how the topological gap could stabilize a Chern insulator at a fractional filling factor near a Kondo breakdown remains to be theoretically analyzed; in particular, why the $v = 4/3$ Chern insulator is only stabilized in the mixed-valence region (region V) but not in the Kondo lattice region (region IV) remains to be better understood. Lastly, the substantially enhanced $R_{xx}$ near the Kondo breakdown (Fig. 1f, 2a, 4a) suggests enhanced quasiparticle scattering and/or density of states near the breakdown. Previous studies[18] have also shown that the breakdown is substantially sharpened at low temperatures. These observations suggest that the Kondo breakdown is a quantum phase transition rather than a smooth crossover, as is further supported by a recent thermoelectric study[20]. Whether this transition is quantum critical or not deserves further experimental investigations. Our results motivate further studies on the emergent topological physics in a chiral Kondo system realized in angle-aligned MoTe$_2$/WSe$_2$.

**Methods**
**Device fabrication**
We applied the standard dry transfer method to fabricate the angle-aligned MoTe$_2$/WSe$_2$ devices, as detailed in Ref. [18-22, 26-28]. The crystal orientations of the WSe$_2$ and MoTe$_2$ monolayers and the stacking order between them were determined by angle-resolved optical second-harmonic generation spectroscopy. The top gate graphite, hBN and monolayer WSe$_2$ and monolayer MoTe$_2$ flakes were picked up sequentially and released to the prepatterned platinum electrodes on hBN/graphite stacks transferred onto SiO$_2$/Si substrates. The MoTe$_2$ exfoliation and the stacking transfer process were performed in a nitrogen-filled glove box to avoid degradation of MoTe$_2$.

**Electrical transport measurements**
The electrical transport measurements were performed in a dilution refrigerator (Bluefors LD250) down to about 20mK of lattice temperature, and also in a top-loading Helium-3 refrigerator with a 20T superconducting magnet. A standard low-frequency (7–17 Hz) lock-in technique was employed to measure the device resistances. A bias current of 1–5 nA was applied by sourcing from a function generator in series with a 10MΩ resistor. Gate voltages were supplied by two Keithley 2400 source meters. A voltage preamplifier with an input impedance of 100MΩ was employed to measure voltages with an enhanced signal-to-noise ratio. The bias current and voltage drop (at the voltage probes) were simultaneously recorded (Stanford Research SR830 and SR860). We calibrated the moiré density and moiré period from the SdH oscillations observed in the $R_{xx}$ map versus $v$ and $E$ at high magnetic fields.

**References**
[1] Kirchner, S. et al. Colloquium: heavy-electron quantum criticality and single-particle spectroscopy. *Rev. Mod. Phys.* **92**, 011002 (2020).
[2] Paschen, S. & Si, Q. Quantum phases driven by strong correlations. *Nat. Rev. Phys.* **3**, 9–26 (2021).
[3] Stewart, G. Heavy-fermion systems. *Rev. Mod. Phys.* **56**, 755–787 (1984).
[4] Vaňo, V. et al. Artificial heavy fermions in a van der Waals heterostructure. *Nature* **599**, 582–586 (2021).
[5] Li, G. et al. Two-dimensional Fermi surfaces in Kondo insulator SmB$_6$. *Science* **346**, 1208–1212 (2014).
[6] Dzero, M. et al. Topological Kondo insulators. *Phys. Rev. Lett.* **104**, 106408 (2010).
[7] Hertz, J. Quantum critical phenomena. *Phys. Rev. B* **14**, 1165 (1976).
[8] Millis, A. Effect of a nonzero temperature on quantum critical points in itinerant fermion systems. *Phys. Rev. B* **48**, 7183 (1993).
[9] Gegenwart, P. et al. Quantum criticality in heavy-fermion metals. *Nat. Phys.* **4**, 186–197 (2008).
[10] Wirth, S. & Steglich, F. Exploring heavy fermions from macroscopic to microscopic length scales. *Nat. Rev. Mater.* **1**, 16051 (2016).
[11] Stewart, G. Non-Fermi-liquid behavior in d- and f-electron metals. *Rev. Mod. Phys.* **73**, 797–855 (2001).
[12] Löhneysen, H. et al. Fermi-liquid instabilities at magnetic quantum phase transitions. *Rev. Mod. Phys.* **79**, 1015–1075 (2007).
[13] Mydosh, J. & Oppeneer, P. Colloquium: Hidden order, superconductivity, and magnetism: the unsolved case of URu$_2$Si$_2$. *Rev. Mod. Phys.* **83**, 1301–1322 (2011).
[14] Allan, M. et al. Imaging Cooper pairing of heavy fermions in CeCoIn$_5$. *Nat. Phys.* **9**, 468–473 (2013).
[15] Jiao, L. et al. Chiral superconductivity in heavy-fermion metal UTe$_2$. *Nature* **579**, 523 (2020).

[16] Simeth, W. et al. A microscopic Kondo lattice model for the heavy fermion antiferromagnet $CeIn_3$. *Nat. Commun.* **14**, 8239 (2023).
[17] Kumar, A. et al. Gate-tunable heavy fermion quantum criticality in a moiré Kondo lattice. *Phys. Rev. B* **106**, L041116 (2022).
[18] Zhao, W. et al. Gate-tunable heavy fermions in a moiré Kondo lattice. *Nature* **616**, 61–65 (2023).
[19] Zhao, W. et al. Emergence of ferromagnetism at the onset of moiré Kondo breakdown. *Nat. Phys.* **20**, 1772–1777 (2024).
[20] Zhang, Y. et al. Thermoelectricity of moiré heavy fermions in $MoTe_2$/$WSe_2$ bilayers. Preprint at https://arxiv.org/abs/2510.26958 (2025).
[21] Zhao, W. et al. Emergence of Chern metal in a moiré Kondo lattice. Preprint at https://arxiv.org/abs/2506.14063 (2025).
[22] Han, Z. et al. Evidence of topological Kondo insulating state in $MoTe_2$/$WSe_2$ moiré bilayers. Preprint at https://arxiv.org/abs/2507.03287 (2025).
[23] Guerci, D. et al. Chiral Kondo lattice in doped $MoTe_2$/$WSe_2$ bilayers. *Sci. Adv.* **9**, eade7701 (2023).
[24] Mendez-Valderrama, J. et al. Correlated topological mixed-valence insulators in moiré heterobilayers. *Phys. Rev. B* **110**, L201105 (2024).
[25] Xie, F. et al. Kondo effect and its destruction in heterobilayer transition metal dichalcogenides. *Phys. Rev. Res.* **6**, 013219 (2024).
[26] Li, T. et al. Continuous Mott transition in semiconductor moiré superlattices. *Nature* **597**, 350–354 (2021).
[27] Li, T. et al. Quantum anomalous Hall effect from intertwined moiré bands. *Nature* **600**, 641–646 (2021).
[28] Zhao, W. et al. Realization of the Haldane Chern insulator in a moiré lattice. *Nat. Phys.* **20**, 275–280 (2024).
[29] Zhang, Y. et al. Spin-textured Chern bands in AB-stacked transition metal dichalcogenide bilayers. *Proc. Natl Acad. Sci. USA* **118**, e2112673118 (2021).
[30] Chang, X. et al. Electric-field-tuned consecutive topological phase transitions between distinct correlated insulators in moiré $MoTe_2$/$WSe_2$ heterobilayer. *Phys. Rev. Lett.* (2026).
[31] Polshyn, H. et al. Topological charge density waves at half-integer filling of a moiré superlattice. *Nat. Phys.* **18**, 42-47 (2021).
[32] Saito, Y. et al. Hofstadter subband ferromagnetism and symmetry-broken Chern insulators in twisted bilayer graphene. *Nat. Phys.* **17**, 478-481 (2021).
[33] Xie, Y. et al. Fractional Chern insulators in magic-angle twisted bilayer graphene. *Nature* **600**, 439 (2021).
[34] He, M. et al. Symmetry-Broken Chern Insulators in Twisted Double Bilayer Graphene. *Nano Lett.* **23**, 11066–11072 (2023).
[35] Su, R. et al. Moiré-driven topological electronic crystals in twisted graphene. *Nature* **637**, 1084–1089 (2025).
[36] Waters, D. et al. Chern Insulators at Integer and Fractional Filling in Moiré Pentalayer Graphene. *Phys. Rev. X* **15**, 011045 (2025).
[37] Zhang, Z. et al. Cascade of Zero-field Chern Insulators in Magic-angle Bilayer Graphene. *Nat. Sci. Rev.* nwaf265 (2025).

[38] Kitagawa, S. et al. Metamagnetic behavior and Kondo breakdown in heavy-fermion CeFePO. *Phys. Rev. Lett.* **107**, 277002 (2011).
[39] Pan, H. et al. Topological Phases in AB-Stacked MoTe$_2$/WSe$_2$: $\mathbb{Z}_2$ Topological Insulators, Chern Insulators, and Topological Charge Density Waves. *Phys. Rev. Lett.* **129**, 056804 (2022).
[40] Cai, J. et al. Signatures of fractional quantum anomalous Hall states in twisted MoTe$_2$. *Nature* **622**, 63 (2023).
[41] Zeng, Y. et al. Thermodynamic evidence of fractional Chern insulator in moiré MoTe$_2$. *Nature* **622**, 69–73 (2023).
[42] Park, H. et al. Observation of fractionally quantized anomalous Hall effect. *Nature* **622**, 74 (2023).
[43] Xu, F. et al. Observation of Integer and Fractional Quantum Anomalous Hall Effects in Twisted Bilayer MoTe$_2$. *Phys. Rev. X* **13**, 031037 (2023).
[44] Lu, Z. et al. Fractional quantum anomalous Hall effect in multilayer graphene. *Nature* **626**, 759 (2024).

## Acknowledgements

We thank Fengcheng Wu and Yang Zhang for their insightful discussions. A portion of this work was carried out at the Synergetic Extreme Condition User Facility (SECUF, https://cstr.cn/31123.02.SECUF).

# Figures

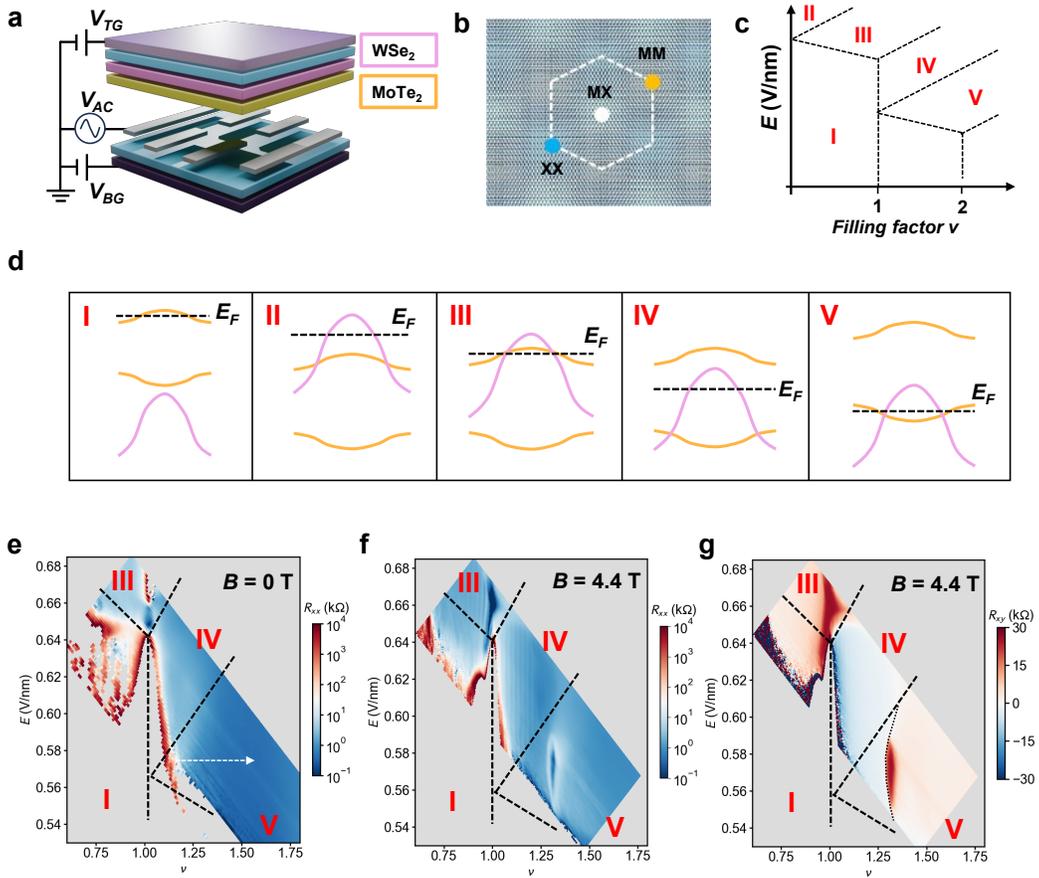

**Figure 1. Electrostatics phase diagram of angle-aligned MoTe₂/WSe₂ bilayers. a**, Schematic dual-gated Hall bar device of an angle-aligned MoTe$_2$/WSe$_2$ moiré bilayer. **b**, Schematic moiré lattice formed by the lattice-mismatched TMD layers, showing the three high-symmetry stacking sites (MM, MX, XX). **c**, Schematic electrostatics phase diagram as a function of the hole filling factor $v$ and the perpendicular electric field $E$. Five distinct regions (I-V) are separated by dashed lines. **d**, Schematic band alignment in regions I-V, where the pink and yellow colors correspond to the WSe$_2$ moiré band and the MoTe$_2$ Hubbard bands, respectively. The Fermi level is denoted by a dashed line. **e,f**, Longitudinal resistance $R_{xx}$ as a function of $v$ and $E$ at perpendicular magnetic fields $B = 0$T (**e**) and 4.4T (**f**). **g**, The corresponding Hall resistance $R_{xy}$ at $B = 4.4$T. The dashed lines mark the boundaries separating the different regions, and the dotted line marks the Kondo breakdown boundary, where $R_{xy}$ changes sign. A pronounced $R_{xy}$ hotspot together with a $R_{xx}$ dip emerges at $v = 4/3$ in the mixed-valence region (region V). All data were taken at a lattice temperature $T = 20$mK.

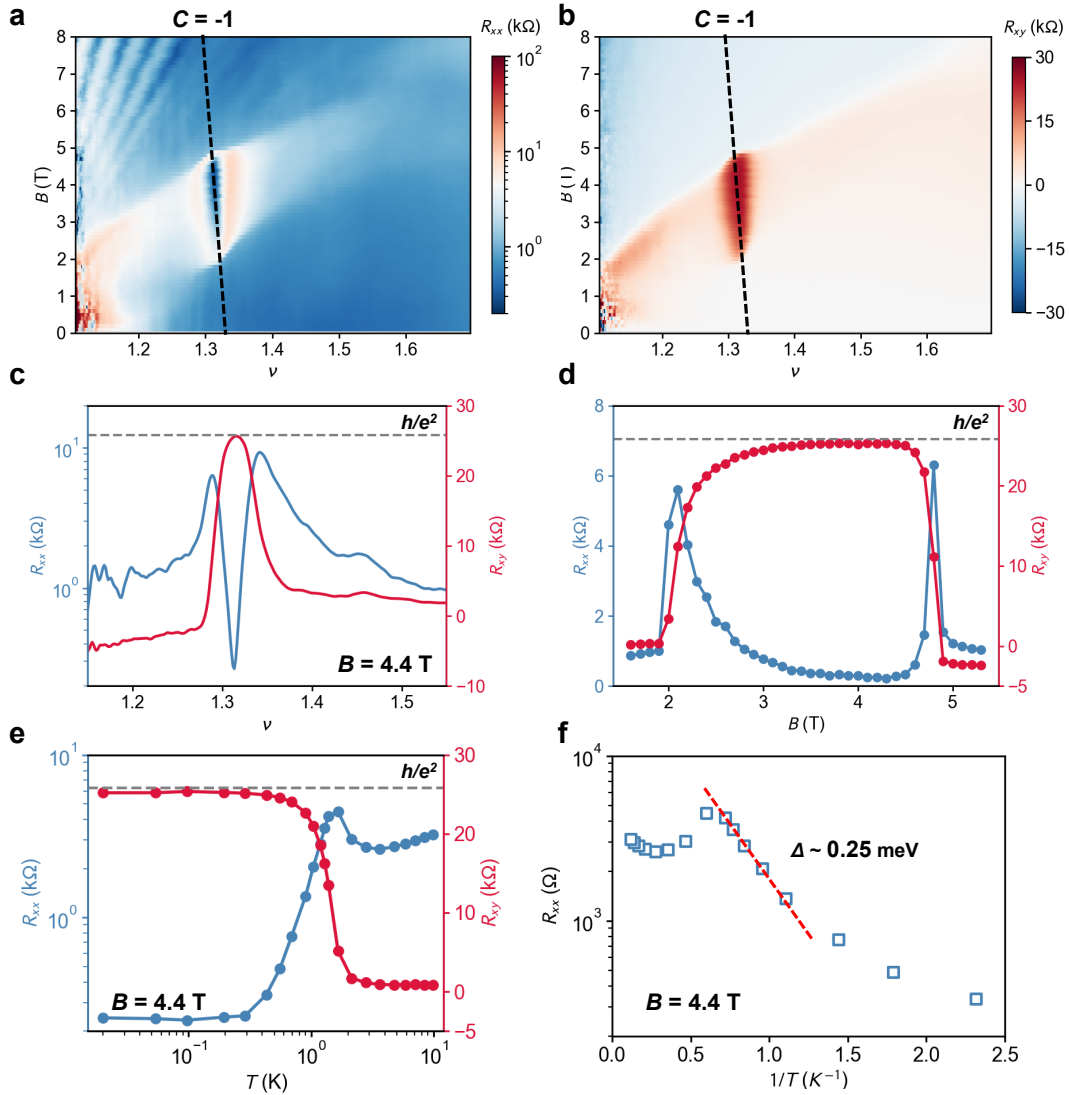

**Figure 2. Symmetry-broken Chern insulator at $\nu = 4/3$.** (**a**) $R_{xx}$ and (**b**) $R_{xy}$ as a function of $\nu$ and $B$ at a fixed electric field $E = 0.575$V/nm (along the dashed arrow in Fig. 1e). The dashed line marks the dispersion of a $C = -1$ Chern insulator based on the Streda formula. **c**, Line cut of $R_{xx}$ (blue) and $R_{xy}$ (red) at $B = 4.4$T, showing a quantized Hall resistance $\frac{h}{e^2}$ and a concomitant $R_{xx}$ dip at $\nu = 4/3$. **d**, $R_{xx}$ and $R_{xy}$ as functions of $B$ along the dashed line in (**a**) and (**b**). The Hall resistance is nearly quantized for $3 \lesssim B \lesssim 5$T and changes sign abruptly at the critical field $B_C \approx 4.7$T. **e**, Temperature dependence of $R_{xx}$ and $R_{xy}$ at $\nu = 4/3$ and $B = 4.4$T. **f**, Arrhenius plot of $R_{xx}$ yielding an energy gap about 0.25meV (3K). The dashed line is a thermal-activation fit.

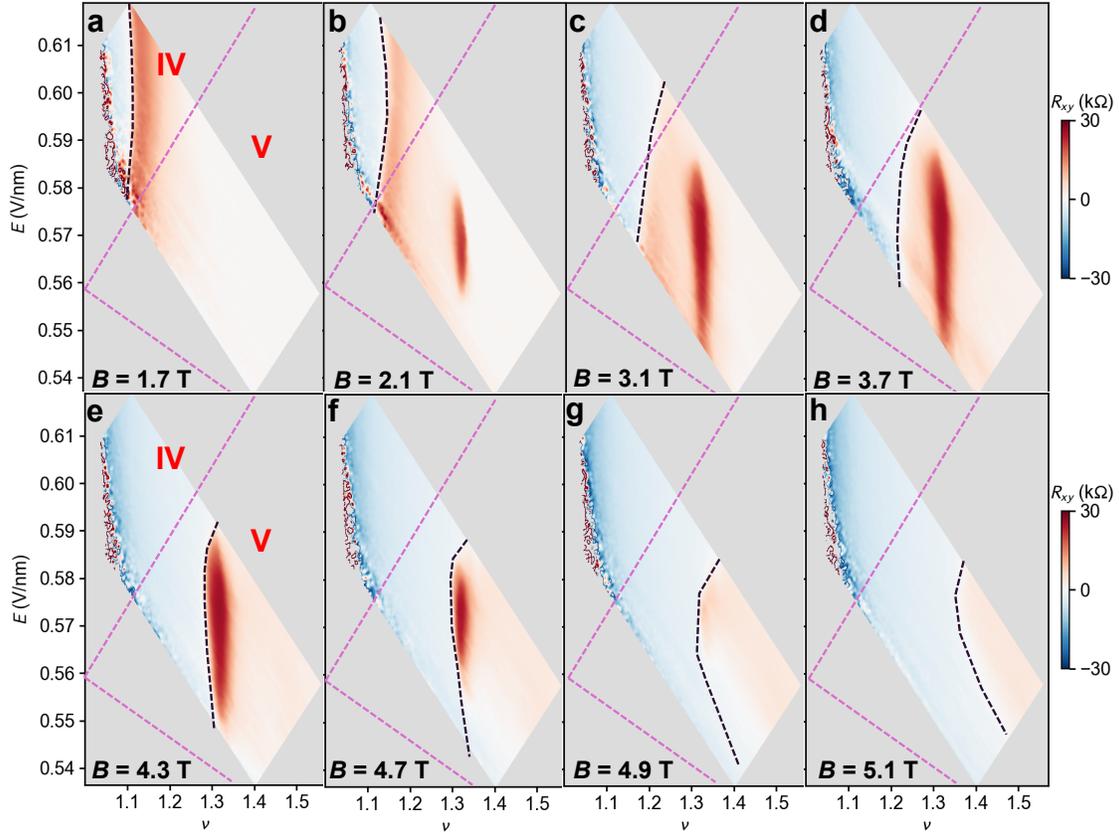

**Figure 3. Evolution of the $\nu = 4/3$ Chern insulator with magnetic field and its connection to Kondo breakdown. a-h**, Series of $R_{xy}$ maps as a function of $\nu$ and $E$ with increasing $B$ (values indicated on each panel). Purple dashed lines represent the phase boundaries separating the different regions in the phase diagram. Dotted lines trace the Kondo breakdown boundary (where $R_{xy}$ changes sign). The boundary extends from the Kondo lattice region (IV) into the mixed-valence region (V) and shifts to higher fillings with increasing field.

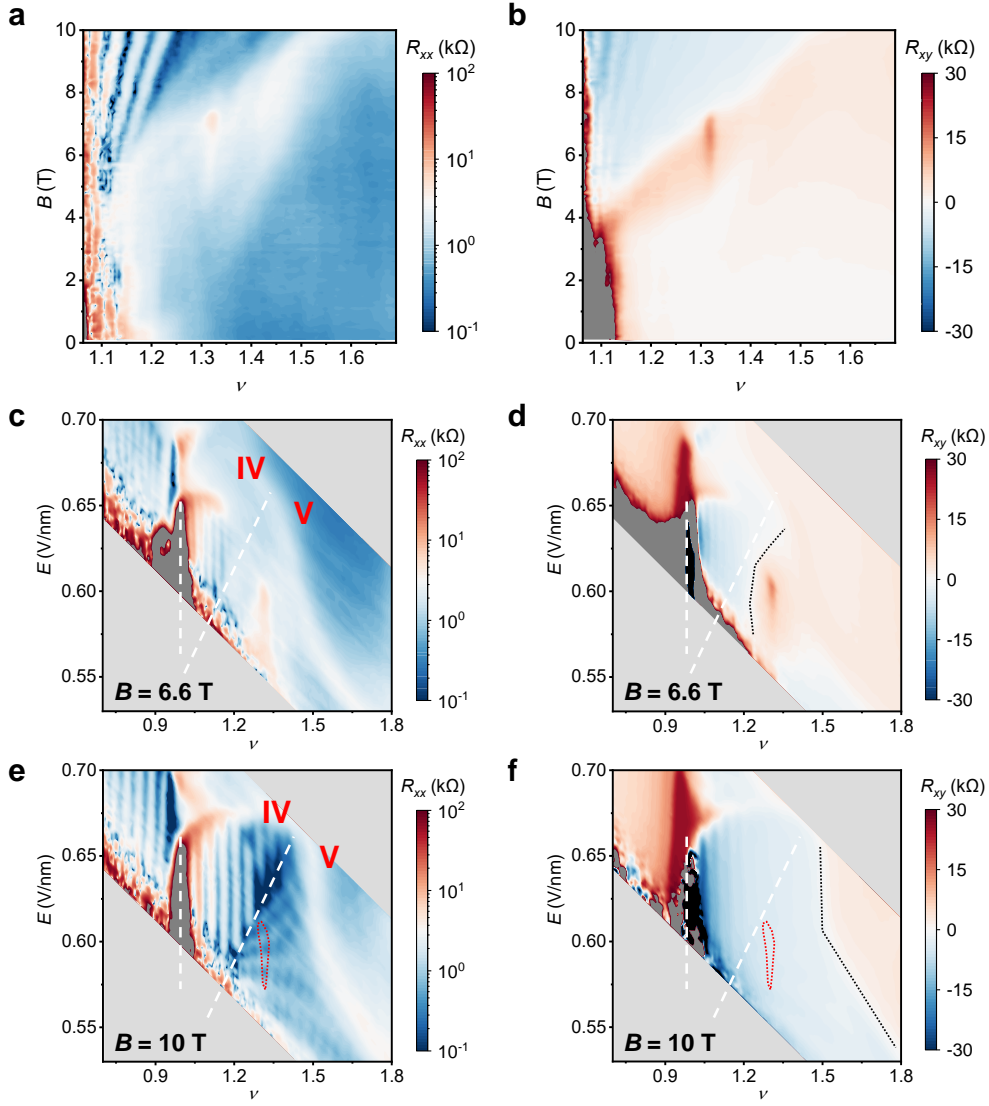

**Figure 4. The $\nu = 4/3$ state in Device II with $\approx 1.5°$ twist angle.** (**a**) $R_{xx}$ and (**b**) $R_{xy}$ as functions of $\nu$ and $B$ at fixed $E = 0.605$V/nm. A $R_{xy}$ hotspot emerges near $\nu = 4/3$ for $5 \lesssim B \lesssim 7$T and disappears suddenly above $B \approx 7$T. **c-f**, $R_{xx}$ (**c,e**) and $R_{xy}$ (**d,f**) versus $\nu$ and $E$ at $B = 6.6$T (**c,d**) and 10T (**e,f**). White dashed lines tentatively mark the phase boundaries for the Kondo lattice (IV) and the mixed-valence (V) regions based on the vertical stripes of SdH oscillations. A $R_{xy}$ hotspot appears near $\nu = 4/3$ at $B = 6.6$T in region V, close to the Kondo breakdown boundary (black dotted line). The hotspot disappears at $B = 10$T, where the Kondo breakdown boundary has moved past $\nu = 4/3$. The red dotted contour in (**e,f**) marks the $R_{xy}$ hotspot location in **d**. All data were measured at $T = 300$mK.

**Supplementary Figures**

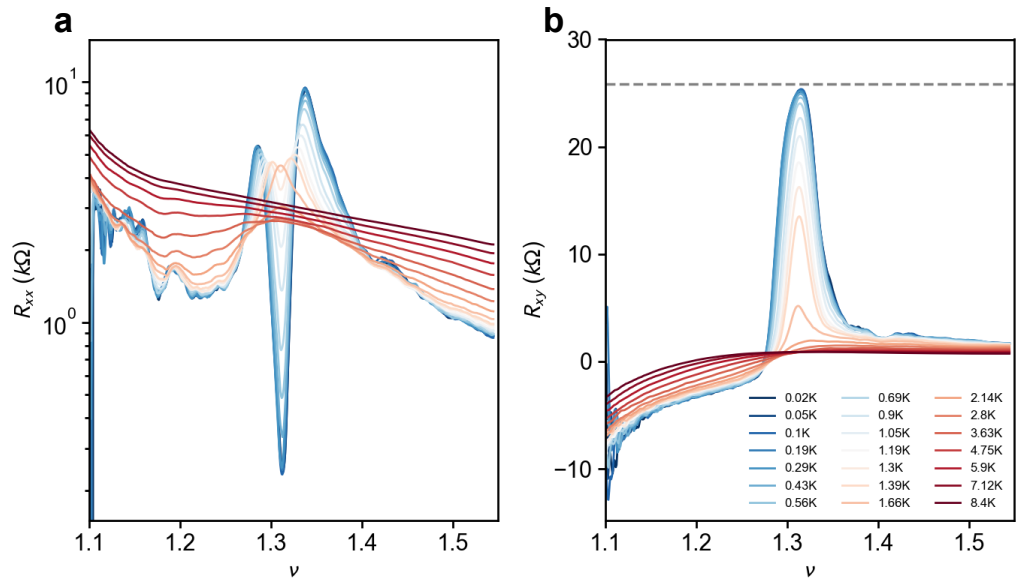

**Supplementary Figure 1. Additional temperature dependence data for Device I.** (a) $R_{xx}$ and (b) $R_{xy}$ as a function of $\nu$, measured at $E = 0.575$V/nm and $B = 4.4$T, from $T = 20$mK to 8.4K. The dashed line marks the quantized Hall resistance at $\frac{h}{e^2}$.

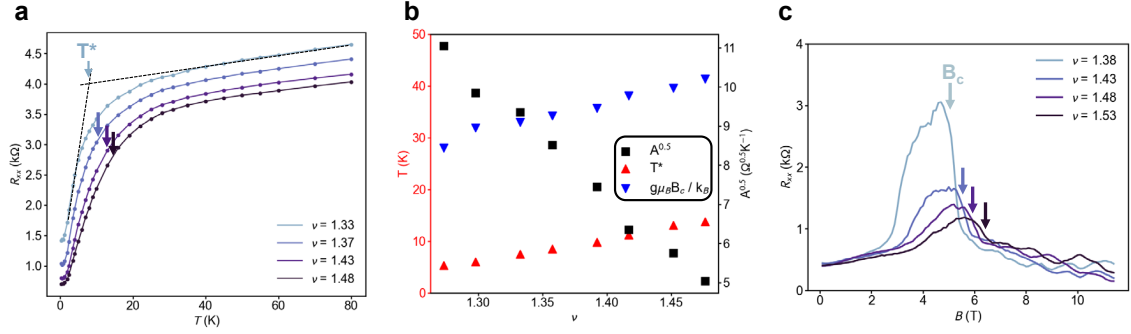

**Supplementary Figure 2. Transport evidence of heavy Fermi physics in the mixed-valence region. a**, Temperature dependence of $R_{xx}$ at $E = 0.575$ V/nm, $B = 0$ T and selected filling factors $\nu$. Dashed lines are linear extrapolations of the data to determine the Kondo coherence temperature $T^*$. **b,** Filling factor dependence of $\sqrt{A}$, $T^*$, and $\frac{g\mu_B B_C}{k_B}$. (Here $g \approx 10$, $\mu_B$ and $k_B$ denote the hole g-factor of TMDs, the Bohr magneton and the Boltzmann constant, respectively.) With increasing $\nu$, the residual resistance ($R_0$) decreases; the Kondo coherence temperature $T^*$ and the critical field $B_C$ increase; and the Kadowaki-Woods coefficient ($A$) decreases. The increase in $T^*$ and the corresponding decrease in $\sqrt{A} \propto (T^*)^{-1}$ are consistent with the expected strengthening of the effective Kondo exchange interaction with $\nu$. **c,** Magnetic field dependence of $R_{xx}$ at selected fillings. Arrows mark the peak position near $B_C$. With increasing B, $R_{xx}$ increases, reaches a peak near the critical field $B_C$ and decreases quickly for $B > B_C$, where SdH oscillations also appear. Note that $R_{xy}$ also changes sign at $B_C$ (Fig. 2a,b). The results in **a-c** show that the phenomenology in the mixed-valence region is similar to that in the Kondo lattice region, suggesting the emergence of a heavy Fermi liquid for $T \lesssim T^*$.

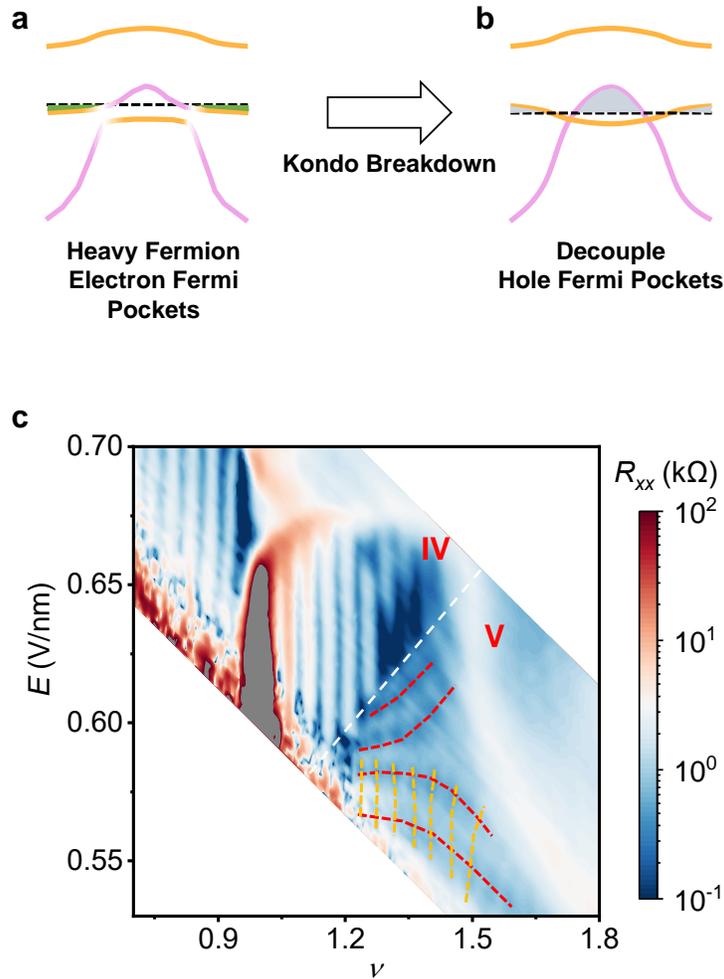

**Supplementary Figure 3. Kondo breakdown in the mixed-valence region. a,b,** Schematic bands and Fermi surfaces for the heavy fermion phase (**a**) and the decoupled phase (**b**) through the magnetic field-induced Kondo breakdown. Dashed lines denote the Fermi level. Below $B_C$, Kondo hybridization produces a heavy Fermi liquid with a large electron-like Fermi surface. Above $B_C$, the Kondo singlets break down, the hybridization gap closes, and the system decouples into two independent hole pockets (one for each layer). **c,** The same as Fig. 4e. Two types of carriers can be identified from the SdH oscillations in the mixed-valence region (V): holes from the Mo-layer (yellow dashed lines) and holes from the W-layer (red dashed lines). Here, the SdH oscillations no longer form vertical stripes because both TMD layers share charges and their relative density changes with the electric field. This allows us to attribute oscillations of different orientations to specific layers: those from the top W-layer (dominantly tuned by the top gate) and those from the bottom Mo-layer (primarily controlled by the bottom gate). The SdH oscillations in the W-layer are also clearer because of the higher hole mobility in the W-layer.

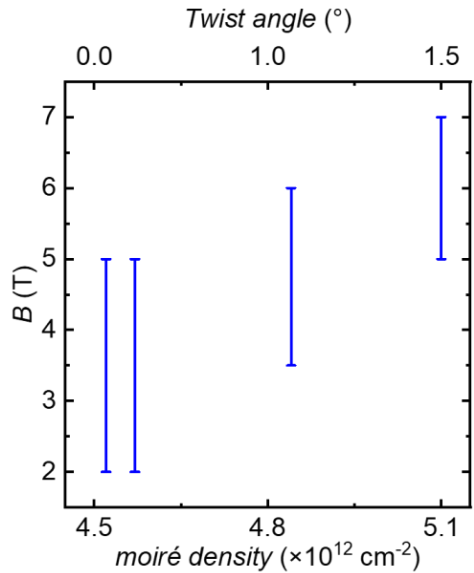

**Supplementary Figure 4. Twisted angle dependence.** Twist angle dependence of the magnetic field range for the appearance of the $v=4/3$ Chern insulator. As the twisted angle (or equivalently, the moiré density) increases, both the onset field and the critical field $B_C$ shift to higher values, while the field range for the $v=4/3$ state narrows.

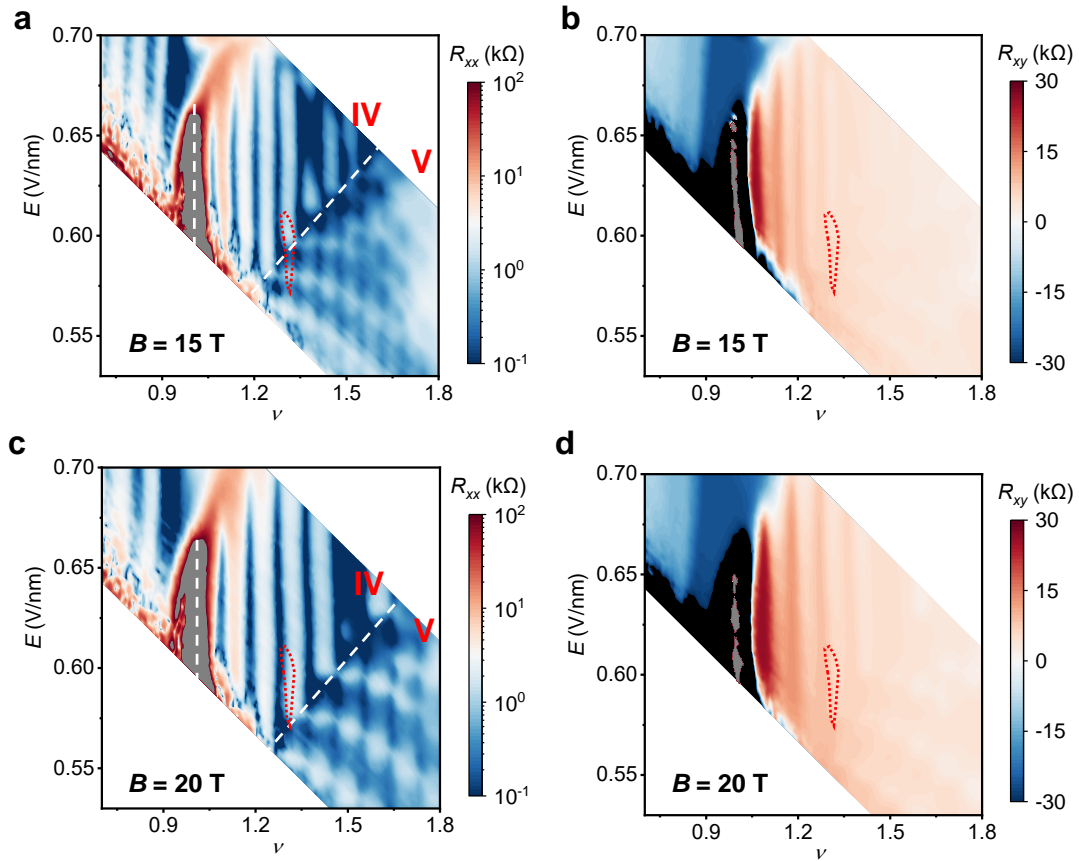

**Supplementary Figure 5.** $R_{xx}$ **and** $R_{xy}$ **maps measured at** $B = $ **15T and 20T (Device II).** The same as Fig. 4c-f except at higher magnetic fields. The boundary separating regions IV and V (white dashed line) shifts towards higher fillings with magnetic field because of the enhanced Mott gap. The red dotted contour marks the $v = 4/3$ $R_{xy}$ hotspot region observed in Fig. 4d at $B = 6.6$T.

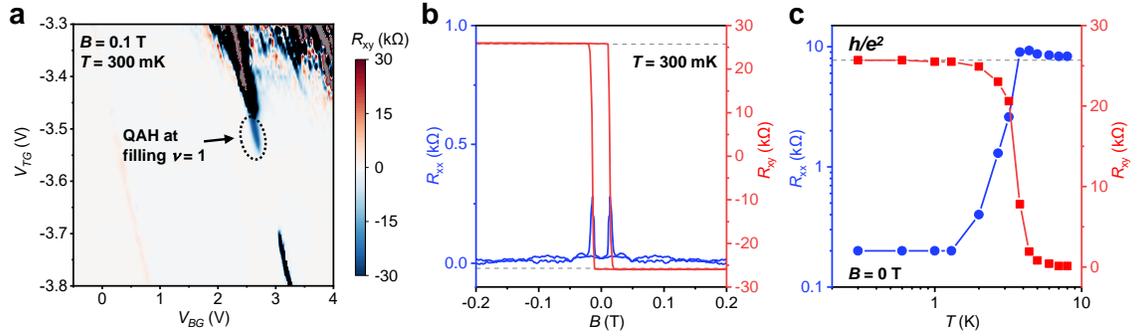

**Supplementary Figure 6. The QAH state at $\nu = 1$ for Device II. a**, The Hall resistance, $R_{xy}$, as a function of the top- and bottom-gate voltages at $B = 0.1$T and $T = 300$mK. The QAH state at $\nu = 1$ is highlighted by the dashed circle. **b,** Magnetic field dependence of $R_{xx}$ and $R_{xy}$ for the QAH state at $T = 300$mK. The dashed lines mark the quantized Hall resistance at $\frac{h}{e^2}$. **c,** Temperature dependence of $R_{xx}$ and $R_{xy}$ for the QAH state at $B = 0$T. $R_{xy}$ becomes quantized at $\frac{h}{e^2}$ below about 2K.